\documentstyle[12pt,twoside]{article}
\newcommand{\ba}{\begin{array}}
\newcommand{\ea}{\end{array}}

\newcommand{\A}{{\cal A}}
\newcommand{\B}{{\cal B}}
\newcommand{\C}{{\cal C}}

\newcommand{\Ha}{{\cal H}}

\newcommand{\M}{{\cal M}}

\def\nz{\ifmmode {I\hskip -3pt N} \else {\hbox {$I\hskip -3pt N$}}\fi}
\def\zz{\ifmmode {Z\hskip -4.8pt Z} \else
       {\hbox {$Z\hskip -4.8pt Z$}}\fi}
\def\qz{\ifmmode {Q\hskip -5.0pt\vrule height6.0pt depth 0pt
       \hskip 6pt} \else {\hbox
       {$Q\hskip -5.0pt\vrule height6.0pt depth 0pt\hskip 6pt$}}\fi}
\def\rz{\ifmmode {I\hskip -3pt R} \else {\hbox {$I\hskip -3pt R$}}\fi}
\def\cz{\ifmmode {C\hskip -4.8pt\vrule height5.8pt\hskip 6.3pt} \else
       {\hbox {$C\hskip -4.8pt\vrule height5.8pt\hskip 6.3pt$}}\fi}

\def\au{{\setbox0=\hbox{\lower1.36775ex%
\hbox{''}\kern-.05em}\dp0=.36775ex\hskip0pt\box0}}
\def\ao{{}\kern-.10em\hbox{``}}

\normalsize
\begin{document}
\bibliographystyle{plain}

\begin{titlepage}
\begin{flushright} UWThPh-2002-28\\

\today
\end{flushright}

\vspace*{2.2cm}
\begin{center}
{\Large \bf The Structure of State Space with Respect to
Imbedding}\\[30pt]

Heide Narnhofer $^\ast $\\ [10pt] {\small\it Institut f\"ur
Theoretische Physik \\ Universit\"at Wien\\ Boltzmanngasse 5,
A-1090 Wien }

\end{center}
\vfill \vspace{0.4cm}

\begin{abstract} The entanglement of formation as well as the
conditional entropy can be used to define leaves in the state
space, given by the linear superposition of their extremal points.
Examples, where these leaves can be found and can be used to
calculate the entanglement respectively the conditional entropy
are presented. The definition of entanglement is generalized to
infinite systems and allows again to find a leaf-structure.
Finally we remark on the additivity property of both expressions,
offering a counter example to the additivity of the conditional
entropy.

\vspace{0.8cm}

\smallskip
Keywords: Entanglement, conditional entropy, symmetry properties
\\
\hspace{1.9cm}

\end{abstract}

\vfill {\footnotesize}

$^\ast$ Heide Narnhofer $^\ast $\\ [10pt] {\small\it Institut
f\"ur Theoretische Physik \\ Universit\"at Wien\\ Boltzmanngasse
5, A-1090 Wien, Tel. +43 1 4277 51516 }E--mail address:
narnh@ap.univie.ac.at
\end{titlepage}
\vfill \eject

\setcounter{page}{2}
\section{Intoduction:}
The phenomenon of entanglement was already well known in the early
stage of quantum mechanics \cite{S 36}. In the near past it has
gained again much interest being a powerful resource in
prospective quantum information techniques. There exist several
expressions to quantify entanglement, depending what features
should be described. One of them is entanglement of formation. It
turns out that this expression not only serves to measure the
costs to produce the entangled state (in the spirit of \cite{BVSW
96},\cite{HHT 01}) but also imposes a structure on the state space
of the composite system, decomposing the state space into
different leaves, a structure that we expect to be useful to
evaluate strategies in quantum encoding \cite{B 96}. Nevertheless
not many examples have been studied so far. In this review we
offer strategies to evaluate such leaves, we collect the known
results and add a few additional ones. We also compare the
structure of state space induced by entanglement with a similar
one, induced by the conditional entropy. This structure is somehow
opposite to the one induced by entanglement, but is not as rigged
and especially does not satisfy additivity with respect to tensor
products, a property that is one of the open questions in the
theory of entanglement.

\section{Entanglement of Formation and the Leaf structure of State
space}

We consider our quantum system to be described by an algebra of
operators $\M $ acting on a Hilbert space $\Ha $. To avoid
topological subtilities we assume in this chapter that the Hilbert
space has finite dimensions. States over $\M $ are given by
density matrices $\rho $ such that $\omega (M)=Tr \rho M.$
Entanglement of formation refers to a subalgebra $\A\subset \M.$

\paragraph{Def. 2.1:}Given a subalgebra $\A\subset \M.$ We define
the entanglement of the state $\omega $ with respect to the
subalgebra $\A $ by $$E(\omega ,\M ,\A )=\inf \sum _i \lambda _i
S(\omega _i )|_{\A }$$ where the infimum is taken over all
possible decompositions $\omega =\sum _i \omega _i $ of the state
$\omega $ into states over $\M .$

\paragraph{Remark:}

i) A special example corresponds to $\M =\A \otimes \B$, where
usually the algebra $\A $ is assigned to Alice and $\B $ to Bob.
In this situation $$E(\omega, \A \otimes \B ,\A )=E(\omega ,\A
\otimes \B ,\B ),$$ but we have also more general imbeddings $\A
\subset \M $ in mind.

ii) The entanglement of formation is a convex function of $\omega
.$ With respect to $\A $ it is monotonically increasing, with
respect to $\M $ it is monotonically decreasing \cite{NT 85}.

iii) Since the Hilbert space is finite the infimum is really
achieved. The $\omega _i $ for which the infimum is achieved are
called optimal decomposers.

The main observation that allows to impose a leaf structure on the
state space and also enables us to evaluate the entanglement for a
larger set of states is the following:

\paragraph{Theorem 2.1:}
Let $f(\omega )$ be a concave function on $\omega $. Let
$$F(\omega )=\inf \sum _i \lambda _i f(\omega _i), \quad \sum _i
\lambda _i \omega _i =\omega .$$ Then the state space $S$
decomposes into leaves $\L ,$ $S =\bigcup \L $ and $F$ is a linear
functional on a leaf $\L $, i.e. $$F(\lambda \omega _1 +(1-\lambda
)\omega _2)=\lambda F(\omega _1 )+(1-\lambda )F(\omega_2 ), \quad
\omega _1 ,\omega _2 \in \L $$ Proof: From concavity it follows
that the infimum is reached at extremal points, i. e. pure states.
Linear decomposition of superpositions of states can only be
better than the linear superpositions of the decompositions, which
makes $F$ convex. Based on this observation different
superpositions of one optimal decomposition can be compared and
this gives the result \cite{BNU 96}.Also we can remember that as a
convex function $F$ is the supremum over affine functionals, and
these affine functionals can be labeled in our situation by $\L $,
$l_{\L} (\omega )=F(\omega )$ for $\omega \in \L $ ,\cite{BNU 02}.

We can collect some properties that these leaves have to satisfy:

i) Let $\alpha _g , g\in G $ be an automorphism group on $\M $
such that $\alpha _g \A \subset \A $ and $\omega \circ \alpha _g
=\omega .$ Let $\overline{\omega } $ belong to the leaf $\L
_{\omega } $. Then also $\overline{\omega }\circ \alpha _g \in \L
_{\omega }.$

ii) We take pure states $\sigma _1,..\sigma _n $ and denote the
corresponding vectors in the Hilbert space $\Ha $ by
$|\sqrt{\sigma _1}\rangle,..., |\sqrt{\sigma _n} \rangle $

\paragraph{Theorem 2.2:} Compatibility relation \cite{BN 01}

We call states $\omega _1$ and $\omega _2 $ compatible if they
belong to the same leaf.

$\sigma _1,..\sigma _n $ are extremal points of the same leaf if
and only if $$ \sum _i |\gamma _i|^2 S(|\sqrt{\sigma _i}\rangle
\langle \sqrt{\sigma _i})|_{\A } + \sum _{i,j}( \gamma _j \gamma
_i^*|\sqrt{\sigma _i}\rangle \langle \sqrt{\sigma _j}|+ \gamma _i
\gamma _j^*|\sqrt{\sigma _j}\rangle \langle \sqrt{\sigma
_i}|)|_{\A }\ln |\sqrt{\sigma _i}\rangle \langle \sqrt{\sigma
_i}|_{\A})$$ $$\leq \langle \sum _i \gamma _i \sqrt{\sigma
_i}|\sum _j \gamma _j\sqrt{\sigma _j}\rangle S(\frac{|\sum _i
\gamma _i \sqrt{\sigma _i}\rangle \langle \sum _j \gamma _j
\sqrt{\sigma _j}}{\langle \sum _i \gamma _i \sqrt{\sigma
 _i}|\sum _j \gamma _j \sqrt{\sigma _j}\rangle}|) $$
for all possible $\gamma _i \in \C.$

The proof can be found in \cite{BN 01}. It is based on
perturbation around the optimal decomposition together with an
application of Theorem 2.1.

As a special case we consider the values $\gamma _1 =1, \gamma _2
= \epsilon,$ all other $\gamma _i =0$. Then we can expand the
inequality. Up to order $\epsilon $ it reduces to the equality
$$Tr |\sqrt{\sigma _1}\rangle \langle \sqrt{\sigma _2}
|(\ln|\sqrt{\sigma _1}\rangle \langle \sqrt{\sigma _1}|-\ln
|\sqrt{\sigma _2 } \rangle \langle \sqrt{\sigma _2}|)=0.$$ Up to
second order in $\epsilon $ an inequality remains, that is not
much more transparent than the general inequality. It is an open
problem whether the above inequality cannot be reduced to a
smaller set of ${\gamma _i}$, e.g. if the compatibility of all
pairs of pure states ($\gamma _i =0$ for all but two elements)
guarantees already that the pure states generate a leaf. So far no
counterexample is known, and in the next chapter we will offer an
example where the leaf is really found on the basis of this
assumption.

\section{Finite dimensional Examples}

A) The simplest example is provided by $\M =M_n\otimes M_k ^0$,
where $M_n$ is a $n$ dimensional full matrix algebra and $M_k ^0$
an abelian algebra of dimension $k.$ Then any state $\omega $ can
be decomposed into $\omega =\sum _{l=1}^k \lambda _l \omega _l $
with $\omega _l$ a pure state on $M_k ^0$. $\omega _l $ can
further be decomposed into pure states over $M_n$. Therefore
$$E(\omega \M, M_n) =E(\omega, \M, M_k ^0)=0$$ and the state space
consists only of one leaf, all pure states being compatible.

B) We take $\M =M_2$ and $\A =M_2^0 =\{\sigma _z\} $ with the
notation of Pauli matrices. Every state over $M_2 $ corresponds to
a density matrix in $M_2$. We choose the special states $\omega $
with $\omega \circ \alpha =\omega $, where $\alpha $ is the
automorphism $\alpha \sigma _z =-\sigma _z, \alpha \sigma _x
=\sigma _x.$ If $|z_1,z_2\rangle $ is an optimal decomposer so is
$|z_2, z_1\rangle.$ To every above $\omega $ we can find an
appropriate pair of such states and can convince us that this pair
satisfies the necessary compatibility relation \cite{BNU 96}.
Therefore the corresponding leaf consists of the orbit under
$\alpha $ of one state and further more the whole state space can
be covered by these leaves after rotation in the $x y$ space.

This example provides us with a possible strategy to search for
optimal decompositions, though it is only applicable if we want to
decompose a state with good symmetry properties.

Assume $\omega \circ \alpha _g =\omega $ $\forall g\in G.$ We look
for a pure state $\overline{\omega }$ such that $\omega = \int
d\eta _g \overline{\omega } \circ \alpha _g,$ i.e. we look for a
state whose orbit under the symmetry group generates the leaf. If
the group is large the orbit might be large too, therefore the
compatibility condition (Theorem 2.2) might be too demanding on
the many $\overline{\omega } \circ \alpha _g.$ Therefore we look
for a subgroup $H\subset G$ such that $\overline{\omega }\circ
\alpha _h=\overline{\omega }$ $\forall h\in H.$ Therefore the
orbit reduces to $G|H$  and should be small enough to satisfy all
compatibility relations but large enough to generate $\omega .$ In
addition we have to be aware that the leaf might be generated by
several orbits. That this strategy can be successful but that all
possibilities we mentioned can be realized will be demonstrated in
the following example:

C) We take $\M =M_3$ and $\A =M_3^0 .$ The relevant group is the
permutation group which is of order $6$. The states $\omega \circ
\alpha _{\pi } =\omega $ are labeled by one parameter, $\omega
(e_{ii} )=1/3,$ $\omega (e _{ij} )=z$ for $i\neq j $ and $-1/6\le
z\le 1/3$ so that the state is positive. For an optimal
decomposition we need at least three states, at most nine.  If
three states are sufficient  the pure state has to be invariant
under a subgroup $H\subset G,$ e.g. without loss of generality we
take the permutations $(2,3).$ This fixes the possible pure state
uniquely depending on $z$. But it turns out \cite{BNU 96} that
this decomposition is not always optimal. We have two bifurcation
points \cite{BNU 02},\cite{TV 00} $-1/6<z_0<0<z_1<1/3.$ This is a
result of numerical analysis but can be made plausible by the
following observation:

For $z=0$ the tracial state can be decomposed into eigenvectors of
$M_3^0$, thus one state corresponds to (1,0,0) and gives
entanglement $E(\omega _0 )=0.$ For $z=1/3$ the state is already
pure and the corresponding vector is $1/\sqrt3(1,1,1)=\psi
_{1/3}.$ For $z=-1/6$ it is easy to find that $\psi _{-1/6}
=1/\sqrt2 (1,-1,0)$ is an optimal decomposer. Here the state is
invariant under the group $H$ but not the vector, only its ray. It
is possible to pass continuously from $\psi _{1/3}$ to $\psi _0$
remaining a fix point of $H$ but not from $\psi _0$ to $\psi
_{-1/6}$. This explains that a bifurcation value has to occur. The
accurate value varies if we vary the concave function $f(\omega )$
in Theorem 2.1 and can therefore not be explained by general
arguments. The bifurcation point $z_1$ is of different nature.
Here we do not break the symmetry of $H$ but we start to need two
orbits with varying weight. This bifurcation point can be found by
a mapping $\Gamma: M_2 \rightarrow M_3;M_2^0 \rightarrow M_3^0$

 $$\left (
\ba{cc} a & c \\ c & d \ea \right ) \rightarrow \left ( \ba{ccc} a
& c/2 & c/2 \\ c/2 & b/2 & b/2 \\ c/2 & b/2 & b/2 \ea \right )
\quad \left (\ba{c} a \\ b \ea \right ) \rightarrow \left ( \ba{c}
a \\ b/2 \\ b/2 \ea \right ). $$ Every decomposition of $\Gamma
(\rho )$ is again into density matrices of the above form and
satisfies especially that $S(\Gamma (\omega _i))|_{M_3^0}$ has the
same monotonicity behaviour with respect to the relevant
parameters as $S(\omega _i)|_{M_2^0}.$ Therefore an optimal
decomposition over $M_2$ can be mapped into an optimal
decomposition over $M_3.$ Especially $$\Gamma 1/\sqrt3 (1,\sqrt2
)=1/\sqrt3 (1,1,1),\quad \Gamma 1/\sqrt3 (\sqrt2 ,1)=1/\sqrt6
(2,1,1).$$ These two vectors combine in the three dimensional case
to a leaf, but they also belong to the leaf that is defined by
$\omega _{1/3}$ respectively to the leaf defined by $\omega
_{z_0}.$ Their orbits under the permutation group generate the
leaf for all $\omega _z ,z_0 \le z\le 1/3$ which can be checked by
comparing with a decomposition of just one orbit for $z_0 <z<1/3.$
This example is in support to the conjecture that a leaf is
determined by the pairs of its extremal points. (Compare the
remark after Theorem 2.2)

\section{Infinite Algebras}

Though in quantum information theory normally one restricts
oneself to finite dimensional algebras it seems worthwhile to
examine how increasing dimensions might influence the structure
and especially whether similar considerations also give some
insight when infinite algebras are imbedded in one another. In
this situation the first problem arises in the definition of the
entanglement, qualitatively and quantitatively, because pure
states on infinite von Neumann algebras do not exist.

A) Let us first consider a simple imbedding: let $\A $ be a type
$II_1$ factor algebra and $\alpha $ a free automorphism (therefore
not an inner automorphism) with $\alpha ^2 =1$ and $\A $ imbedded
into the algebra $\M =\A \bowtie _{\alpha } Z^2$, i.e. the crossed
product of the algebra $\A $ with the automorphism $\alpha $ and
by the assumptions again a type $II_1$ factor. A physical
realization is given with $\M $ the algebra of infinitely many
fermions and $\A $ the subalgebra of even polynomials in creation
and annihilation operators where $\alpha $ is induced by some
$(a_0 +a_0^*).$ We can write elements of $\M $ respectively of $\A
$ conveniently as  $$M= \left ( \ba{cc} A_1 & A_2 \\\alpha A_2  &
\alpha A_1 \ea \right )$$ where $A_1,A_2 $ belongs to $\A $ and
$\A $ is imbedded into $\M $ by demanding that $A_2 =0.$ On $\M $
we can define an automorphism $\hat{\alpha }$ $$\hat{\alpha }M=
\hat{\alpha }\left ( \ba{cc} A_1 & A_2
\\ \alpha A_2 & \alpha A_1 \ea \right )=\left ( \ba{cc} A_1 & - A_2
\\ -\alpha A_2 & \alpha A_1 \ea \right )$$ so that $\A $ is the fix
point algebra under $\hat{\alpha }.$ Notice that the automorphism
$\alpha $ can now be implemented by either of the operators $$
\left ( \ba{cc} V & 0
\\ 0 & V \ea \right ) or \left ( \ba{cc} 0 & 1
\\ 1 & 0 \ea \right )$$ where for the first operator $V\notin \A $
 but the later operator belongs to $\M .$

 To find a definition for the entanglement let us recall the
definitions in the finite case: the entropy itself can be written
[NT 85] as $$S(\omega )= \sup \sum _k \lambda _k S(\omega |\omega
_k ),\quad \omega = \sum _k \lambda _k \omega _k $$ where the
supremum is taken over all possible decomposition and is reached
for every decomposition into pure states. The entanglement then
reads $$E(\omega, \M ,\A )=\inf \sum _i^{\M} \mu _i \sup \sum
_k^{\A } \lambda _{ki} S(\omega_i |\omega _{ik}).$$ Here $\omega
=\sum \mu _i \omega _i $ is decomposed into states over $\M $
whereas $\omega _i =\sum \lambda _{ik} \omega _{ik} $ is
decomposed into states over $\A .$ Every decomposition results
from a positive operator in the relative commutant of a
representation in which the state is given as expectation value
with a vector: $$\omega (A)=\langle \Omega|\Pi(A)|\Omega \rangle
\quad \omega _k (A) =\langle \Omega |Q_k \Pi (A) |\Omega \rangle
$$ where $Q_k \in \Pi(\A )',Q_k\ge 0.$ We can now replace the
definition of the entanglement by $$E(\omega, \M ,\A )= \inf
\sum_i^{\M } \mu _i \sup \sum _k^{\A} \inf _{E _i} \omega _i (Q_k
)S(\omega _i(E_i(Q_k ) \cdot )|\omega _i(Q_k \cdot))|_{\A }$$
where we stay in a common representation for all $\omega _i .$
Here $E_i(Q_k )$ is an $\omega _i $ preserving completely positive
map from $\Pi (A )'$ into $\Pi (\M )'$ and the supremum is taken
over all decompositions $\sum _k Q_k =1$ of operators $Q_k >0 \in
\Pi (\A )'.$  Therefore $Q_k $ contributes to the entanglement
only as far as it is a refinement of a decomposition into states
over $\M .$ Since the infimum is still achieved if $\omega _i $ is
pure over $\M $ (if $\M $ is finite dimensional so that this
statement makes sense) and then $\omega _i(E_i(Q_k ) \cdot
)=\omega _i(Q_k ) \omega _i(\cdot )$ the two definitions coincide
in the finite dimensional case. Especially also in this form
Theorem 2.1 can be applied. But in the infinite case it enables us
to stop with a decomposition into $\omega _i$ already at an early
stage as we will see in the following examples. First we note

\paragraph{ Lemma:} For the algebras $\M =\A \bowtie _{\alpha } Z^2 \supset \A $
 the entanglement of any state $\hat{\omega }$
 satisfies $E(\hat{\omega } ,\M ,\A )\le \ln 2.$

Proof: Let $A'$ be an operator in the relative commutant $\pi (\A
)'$ in the GNS representation induced by the state $\omega $ over
$\A $, where we assume that $\omega $ is faithful, i.e. $\omega
(A)>0 $ for all positive operators $A\in \A $. Then for any
extension $\hat{\omega }$ of $\omega $ as state over $ \M $ we can
write the elements of $\Pi (\M )'$ respectively of $\Pi(\A
)'\supset \Pi (\M )'$ as
 $$\left (
\ba{cc} A_1' & A_2'V \\ A_2'V & A_1' \ea \right )\in \Pi (\M
)'\quad \left ( \ba{cc} A_1' & A_2'V \\ A_3'V & A_4' \ea \right
)\in \Pi (\A )'$$ where the automorphism $\alpha $ is implemented
by $\alpha A =VAV.$ On $\Pi(\A )'$ there exists the automorphism
$$\overline{\alpha } \left ( \ba{cc} A_1' & A_2'V
\\ A_3'V & A_4' \ea \right )= \left ( \ba{cc} A_4' & A_3'V \\
A_2'V & A_1' \ea \right )$$ such that
$E(A')=\frac{1+\overline{\alpha }}{2}A'\in \Pi(\M )'$ is a
conditional expectation from $(\Pi (\A )'$ into $\Pi (\M )'$ that
satisfies $E(Q)\geq \frac{1}{2}Q.$ Since every state over $\M $
$\hat{\omega } $ can be written in the form $$\left \langle
\ba{ccc} \Omega | & .. & |\Omega  \\ \Psi | & .. & |\Psi \ea
\right \rangle + \left \langle \ba{ccc} V\Psi | & .. & |V\Psi
\\ V\Omega | & .. & |V\Omega \ea \right \rangle .$$ It follows that
with $$\left ( \ba{cc} 0 & V \\ V & 0 \ea \right ) |\ba{c} V\Psi
\\ V\Omega \ea  \rangle = |\ba{c} \Omega
\\ \Psi \ea  \rangle$$
the state $\hat{\omega }$ corresponds to a state over $\Pi (\A )'$
for which $\hat{\omega } \circ \overline{\alpha }=\hat{\omega }$
and therefore $\hat{\omega } (E(Q))=\hat{\omega }(Q).$ Together
with the general estimate on the relative entropy that $S(\omega
|\phi )<0$ if $\omega >\phi $ this proves the lemma. We want to
calculate the entanglement for special states and to find the
corresponding leaf.

a) Let $\hat{\omega }(M)=\langle \hat{\Omega }|M|\hat{\Omega
}\rangle $ satisfy $\hat{\omega }\circ \hat{\alpha }=\hat{\omega
}$ i.e. we consider gauge invariant states over $\M .$ All these
states belong to the same leaf and satisfy  $E(\hat{\omega }, \M
,\A )=0.$

Proof: The set of these states is stable under linear
superposition. Further with $$\hat {\omega } (A) =\left \langle
\ba{c} \Omega \\ 0 \ea \left |  \left ( \ba{cc} A_1 &  0 \\ 0 &
\alpha A_1 \ea \right ) \right | \ba{c} \Omega \\ 0 \ea \right
\rangle \quad E(Q) = E\left ( \ba{cc} A_1' & A_2'V \\ A_3'V & A_4'
\ea \right ) = \left( \ba{cc} A_1' & 0 \\ 0 & A_1' \ea \right ) $$
$$\hat{\omega }(E(Q)A)=\hat{\omega }(QA)$$ so that decompositions
by projectors from $\Pi (\A )'$ reduce to decompositions already
in $\M $.

b) Consider states of the form $$\hat{\omega }(M) =\left \langle
\ba{ccc} \Psi | & .. & |\Psi  \\ \Psi | & .. & |\Psi \ea \right
\rangle .$$ All states of this form belong to the same leaf and
for them $E(\hat{\omega } ,\M ,\A )=\ln 2.$

Proof: Every vector that is dominated by $\hat{\omega }$ can be
represented by a vector obtained by the application of some vector
from $\Pi (\M )',$ $\tilde{\omega } (M)=\tilde{\omega } (M'^* M
M')$ where therefore $\tilde{\omega }$ is now implemented by the
vector $$\left (\ba{cc} A_1' & A_2'V \\ A_2'V & A_1' \ea \right )
\left |\ba{c} \Psi \\ \Psi \ea \right  \rangle = \left |\ba{c}
(A_1'+A_2'V)\Psi
\\ (A_1'+A_2'V)\Psi \ea \right \rangle $$
and is therefore of the desired form. The lemma follows if for all
these states we can find an appropriate decomposition such that
for all $E$ $$ \sum _k \hat{\omega }(Q_k) S(\hat{\omega
}(E(Q_k)\cdot |\hat{\omega }(Q_k\cdot )=\ln 2.$$ Let us assume
that $|\Psi \rangle =C'|\Omega \rangle $ for some $C'\in \A' $ and
$|\Omega \rangle $ is the vector implementing the tracial state on
$\A .$ Take   a projection in $\A' $, $P'$ with $\alpha P' =1-P'$
and $[P',C']=0$. Such a projection can be found fore a dense set
of $C'.$ Then for any $E$ $$E \left ( \ba{cc} P' & 0 \\ 0 & 1-P'
\ea \right ) =\left ( \ba{cc} \overline {P'} & 0 \\ 0 &
\overline{P'} \ea \right ) $$ for some projector $\overline{P'}.$
$$\langle \ba{c} \Omega
\\ \Omega \ea | \left ( \ba{cc} C'^* & 0 \\ 0 & C'^* \ea \right )
\left ( \ba{cc} A_1 & 0
\\ 0 & \alpha A_1  \ea \right ) \left ( \ba{cc} P' & 0 \\ 0 & 1-P'
\ea \right ) \left ( \ba{cc} C' & 0 \\ 0 & C' \ea \right )| \ba{c}
\Omega \\ \Omega \ea  \rangle =$$ $$\langle \ba{c} \Omega \\
\Omega \ea | \left ( \ba{cc} C'^*C' & 0 \\ 0 & C'^*C' \ea \right )
| \ba{c} \Omega \\ \Omega \ea  \rangle \langle \omega
|A_1(P'+\alpha (1-P')|\omega \rangle =$$ $$=c\langle \Omega |A_1
P'|\Omega \rangle =\omega _1 (A_1)$$ for appropriately chosen
operators $A_1,P'$ that cluster with $C'$ whereas $\hat{\omega }
(E(Q)A)$ as state over $\A $ is $\alpha $ invariant. Based on the
Kosaki formula for the relative entropy with appropriate variation
on $A_1,P'$ the decomposition by $P'$ is as powerful as the
decomposition of the tracial state into a pure state for a two
dimensional matrix algebra and we can achieve the maximal value
$\ln 2.$

c) We consider the state $\hat{\omega }_U$ induced by the vectors
$ | \ba{c} U\Psi \\ \alpha U \Psi \ea  \rangle $ with $\alpha
U\neq U.$ These states belong to a leaf $\L _U$ on which again $
E(\hat{\omega} _U,\M ,\A )=\ln 2 .$ The leaves $\L _U \neq \L_1.$
Proof: The leaf $\L _U$ results from the automorphism $\gamma _U $
implemented by $\left ( \ba{cc} U & 0 \\ 0 & \alpha U \ea \right )
$ that satisfies $\gamma _U \A \subset \A $ and therefore also
acts as map between leaves. The leaves have to be different
because a linear superposition   of two states of different leaves
dominates a state with vanishing entanglement $$\hat{\omega }_U
+\hat{\omega }_1 \geq c_U \hat{\omega } (1+\hat{\alpha } ) .$$
Consider the states in the leaves that are obtained from the
tracial state by operators from $\Pi (\M )',$ $\left ( \ba{cc} 1 &
1 \\ 1 & 1 \ea \right )+\left ( \ba{cc} 1 & U'VU'^* \\ U'VU'^* & 1
\ea \right ).$ In the spectral representation taking into account
that $U'VU'^*\neq 1$ is selfadjoint and unitary $\left ( \ba{cc} 2
& 1\pm 1
\\ 1\pm & 2 \ea \right )$ we see that in some subspace it acts as the identity
and cannot break the invariance of the initial state under $
\hat{\alpha } .$

This does not implement that $\L _U$ and $\L _1 $ have trivial
intersection, e.g. we can imagine there exists a $\Psi $ such that
$\Psi $ and $U\Psi $ are orthogonal for all $U.$

Collecting the results for the imbedding $\A \subset \M =\A
\bowtie _{\alpha } Z^2$ we notice that the amount of entanglement
varies as for $M_2^0 \subset M_2.$ But to every value of
entanglement there belong infinitely many different leaves
reflecting the size of the algebra.

B) As a completely different example we can consider the imbedding
$\A \subset \M =\A \otimes \B $ where both algebras $\A $ and $\B
$ are infinite algebras. Here we have not succeeded to find a
closed expression for the entanglement. We can only define
$$E(\omega ,\A \otimes \B ,\A )=\sup _n E(\omega ,\A _n \otimes \B
_n ,\A _n ) $$ where $\A _n $ and $\B _n $ are finite dimensional
subalgebras \cite{N 02}. The supremum can be replaced by taking
the limit over any sequence of increasing algebras as a
consequence of the monotonicity properties of the entanglement .
(Compare \cite{N 02} with a more detailed analysis.)

\section{The Conditional Entropy}

Another quantitiy that behaves differently in quantum theory than
in classical theory is the conditional entropy. In classical
theory it is is defined by $$H_{\omega }(\M |\A )=S(\omega )|_{\M
} -S(\omega )|_{\A }$$ which can be generalized to $$H_{\omega
}(\B |\A )=S(\omega )|_{\B \bigvee \A } -S(\omega )|_{\A }$$ if we
do not consider imbeddings. This expression does not work in
quantum theory, on one hand by lack of monotonicity of the
entropy, on the other hand because the algebra $\B \bigvee \A$
generated by the two subalgebras will in general be too big. As a
useful replacement one considers [OP 93] $$H_{\omega }(\B |\A
)=\sup \sum _i \lambda _i [S(\omega |\omega _i)|_{\B } -S(\omega
|\omega _i)|_{\A }]$$ where the supremum is taken over all
possible decompositions $\omega = \sum _i \lambda _i \omega _i $
into states over $\M $ or $\A \bigvee \B .$ Different from
classical theory we can find states for which $$ H_{\omega }(\A
\otimes \B |\A )>H_{\omega }(\B |\A ).$$ The optimal decomposition
for $H_{\omega }(\B |\A )$ asks for a delicate balance not to be
too fine for $\A $ but sufficiently fine for $\B .$ If however we
concentrate on imbeddings $\A \subset \M $ then $H_{\omega }(\M
|\A )$ has some analogies with the entanglement.

 With$$H_{\omega }(\M |\A )=\sup \sum _i \lambda
_i [S(\omega |\omega _i)|_{\M } -S(\omega |\omega _i)|_{\A }]$$
the conditional entropy is concave in $\omega $ and the supremum
is achieved for pure states $\omega _i.$ This can be seen by the
following observations:
 Refinement of the decomposition improves the estimate because $$
\sum _i \sum _j  \lambda _{ij} S(\omega |\omega _{ij} ) =\sum _i
\sum _j \lambda _{ij} S(\omega |\omega _i )+\sum _i \sum _j
\lambda _{ij} S(\omega _i |\omega _{ij} )$$ and $$S(\omega
_i|\omega _{ij})|_{\M } -S(\omega _i|\omega _{ij})|_{\A }\geq 0 $$
for $\M \supset \A .$ For pure states $\omega _i $ $$\sum _i
\lambda _i[S(\omega |\omega _i)|_{\M } -S(\omega |\omega _i )|_{\A
} )=S(\omega )|_{\M } -S(\omega )|_{\A } + \sum \lambda _i
S(\omega _i )|_{\A }.$$ For the last expression we have to look
for the supremum instead of looking for the infimum as we did for
calculating the entanglement. We can apply a variational principle
([KW 95]) that is conclusive as long as we do not reach   the
boundary of the area of permitted decompositions.  This boundary
will not be reached if we limit the number of states in the
decomposition sufficiently. The variation of the entropy defines a
vector valued function (compare also [BNU 96])$$F(|\phi >) =\frac
{\partial}{\partial<\psi|} ||\psi ||^2 S(\frac{|\phi +\psi
><\phi +\psi |}{<\phi +\psi |\phi +\psi >})$$ that satisfies $F(c|\phi>)=cF(|\phi>).$
Together with the condition $\rho =\sum _i \lambda _i|\phi_i><\phi
_i|$ this reduces to a kind of eigenvalue equation $$F(|\phi
_i>)+M(\rho )|\phi _i>=0$$ with $M(\rho )$ acting as Lagrange
multiplier. Due to linearity it follows that with $\sum _i\lambda
_i|\phi _i><\phi _i|$ being an optimal decomposition also $\sum
_i\mu _i|\phi _i><\phi _i|$ is an extremal decomposition for some
$\overline{\rho },$ i.e. $M(\rho )=M(\overline{\rho } )$ serves as
Lagrange multiplier also for the new $\overline{\rho }.$ Of course
we have to keep the possibility in mind that a supremum might
change into a saddle point. Apart from this restriction we can
conclude that if a set $(\omega _i)$ is optimal with respect to
$\omega =\sum _i \lambda _i \omega _i $ then it is also optimal
with respect to $\overline{\omega }=\sum _i \mu _i \omega _i .$ In
this situation the conditional entropy also defines leaves in the
state space.

 If we look for a compatibility condition similar as
for the entanglement then it just turns into the opposite
inequality
 $$ \sum _i |\gamma _i|^2 S(|\sqrt{\sigma _i}\rangle
\langle \sqrt{\sigma _i})|_{\A } + \sum _{i,j}( \gamma _i \gamma
_j^*|\sqrt{\sigma _i}\rangle \langle \sqrt{\sigma _j}|+ \gamma _j
\gamma _i^*|\sqrt{\sigma _j}\rangle \langle \sqrt{\sigma _i}|)_{\A
}\ln |\sqrt{\sigma _i}\rangle \langle \sqrt{\sigma _i}|_{\A})$$
$$\geq \langle \sum _i \gamma _i \sqrt{\sigma _i}|\sum _j \gamma
_j\sqrt{\sigma _j}\rangle S(\frac{|\sum _i \gamma _i \sqrt{\sigma
_i}\rangle \langle \sum _j \gamma _j \sqrt{\sigma _j}}{\langle
\sum _i \gamma _i \sqrt{\sigma
 _i}|\sum _j \gamma _j \sqrt{\sigma _j}\rangle}) $$
 now with the restriction that may be the inequality only holds for a
 restricted area of $\gamma _i.$

 The similarity of the compatibility relation is of interest in
 the context of one of the open problems in the theory of
 entanglement: is the entanglement additive, i.e. is $$E(\omega
 _1\otimes \omega _2 ,\M _1 \otimes \M _2 , \A _1 \otimes \A_2 )
=E(\omega _1 ,\M _1 ,\A _1 )+E(\omega _2 ,\M _2 ,\A _2 )?$$ Known
examples support the conjecture. Also if $E(\omega _2 ,\M _2 ,\A
_2 )=0$ then equality follows from $$E(\omega
 _1\otimes \omega _2 ,\A _1 \otimes \B _1 \otimes \A_2 \otimes \B_2 , \A _1 \otimes \A_2 )
\geq E(\omega
 _1\otimes \omega _2 ,\A _1\otimes \B_1 \otimes \A _2 , \A _1 \otimes \A_2
 )=$$ $$E(\omega
 _1\otimes \omega _2 ,\A _1 \otimes \B _1 \otimes \A _2 , \B _1
 )\geq E(\omega
 _1\otimes \omega _2 ,\A _1 \otimes \B _1 , \B _1 )=E(\omega
 _1\otimes \omega _2 ,\A _1 \otimes \B _1 , \A _1 ).$$
In a more general situation the additivity of entanglement
translated to the leaf structure of tensor products would demand
that with $(\sigma _1, \sigma_2 )$ belonging to a leaf of one part
and $(\rho _1 ,\rho _2 ) $ belonging to a leaf of the other then
$\sigma _1 \otimes \rho _1 $ and $\sigma _2 \otimes \rho _2 $ have
to belong to the same leaf in the tensor product. In the
inequality $S(|\sum \gamma _i \sqrt{\sigma _i} \otimes \sqrt{\rho
_i}>)$ is the only term that does not factorize and has to be
estimated on the basis of $S(|\sum _i \gamma _i'\sqrt{\sigma
_i}>)$ and $S(|\sum _i \gamma _i''\sqrt{\rho _i}>).$ Such an
estimate is missing so far. But it can support additivity either
for the entanglement or for the conditional entropy. But for the
conditional entropy we will give already a counter example to
additivity. This example shows that provided some relation between
the entropies above exist then it can only support additivity of
the entanglement.
\paragraph{Example:} Consider the tracial state on $\A \otimes \B
\otimes \C $ with $\A  =M_{n^2},\B =M_n,\C =M_n.$ Then $$H_{\tau
}(\A \otimes \B \otimes \C |\B \otimes \C )=4\ln n,\quad H_{\tau }
(\A \otimes \B |\B )=2\ln n. $$ whereas with $ H_{\tau } (\C |\C
)=0$ additivity would demand identity of the two expressions.

 At last we present a simple example where the conditional
entropy can be calculated based on similar considerations as for
the entanglement and really gives a leaf structure in the state
space that is in some sense opposite to the one defined by the
entanglement:

\paragraph{Example:} Consider $M _n \supset M_n^0 =[P_i,i=1..n].$
Take $\rho =\lambda _iP_i.$ This state is invariant under unitary
transformations $U\in M_n^0.$  Therefore we can take $M_n^0=G$,
the group under consideration that generates the orbit in the
leaf. Take $Q$ a one dimensional projector that satisfies $Tr QP_i
=1/n$, e.g. $Q=1/ n|1,...1><1..1|.$ The orbit of $Q$ defines a
complete set of vectors in the Hilbert space and we can pick
$Q_1,..Q_n$ with $\sum _i Q_i =1.$ Therefore $\sqrt{\rho }Q_i
\sqrt{\rho } $ decomposes $\rho $ and satisfies
$$S(\frac{\sqrt{\rho }Q_i \sqrt{\rho }}{Tr\rho Q_i }) =S(\rho ).$$
Taking into account the concavity of the entropy we have therefore
achieved the optimal decomposition and $$H_{\omega
}(M_n|M_n^0)=S(\omega )|_{M_n}.$$

\end{document}